\documentclass[aps,pre,reprint,superscriptaddress,nofootinbib]{revtex4-2}

\usepackage[T1]{fontenc}
\usepackage[utf8]{inputenc}
\usepackage{amsmath,amssymb,bm}
\usepackage{graphicx}
\usepackage{microtype}
\usepackage[hidelinks]{hyperref}
\usepackage{placeins}
\newcommand{\DTV}{D_{\mathrm{TV}}}

\begin{document}

\title{Superradiant Mpemba Relaxation in a Dicke Ladder}

\author{Matheus G. H. Santos}
\affiliation{Instituto de Física, Universidade Federal de Goiás,
74690-900 Goiânia, Goiás, Brazil}

\author{Hugo Sanchez}
\affiliation{Instituto de Física de São Carlos, Universidade de São Paulo, 135560-970, São Carlos-SP, Brazil}

\author{Italo M. de Araújo}
\affiliation{Instituto de Física de São Carlos, Universidade de São Paulo, 135560-970, São Carlos-SP, Brazil}

\author{Luis Felipe A. da Silva}
\affiliation{Instituto de Física de São Carlos, Universidade de São Paulo, 135560-970, São Carlos-SP, Brazil}

\author{Miled H. Y. Moussa}
\affiliation{Instituto de Física de São Carlos, Universidade de São Paulo, 135560-970, São Carlos-SP, Brazil}

\author{Norton G. de Almeida}
\email{norton@ufg.br}
\affiliation{Instituto de Física, Universidade Federal de Goiás,
74690-900 Goiânia, Goiás, Brazil}

\date{\today}

\begin{abstract}
The Mpemba effect occurs when a state initially farther from stationarity
overtakes a closer one during relaxation. We show that this anomalous ordering
can coexist with superradiant emission in a collectively damped ensemble of
two-level systems. In the symmetric Dicke manifold, zero-temperature decay is
a population cascade with rates $\Gamma n(N-n+1)$. We construct the sparse
family $\rho_A=|\lceil2N/3\rceil\rangle\langle\lceil2N/3\rceil|$ and
$\rho_B=(|0\rangle\langle0|+|N\rangle\langle N|)/2$. Although $A$ is initially
more energetic and more distant from the stationary ground state, it crosses
$B$ in both energy and trace distance because it starts in a more radiative
region of the Dicke ladder. Both states develop emission peaks larger than
their independent-emitter references. The crossing persists for all examined
sizes from $N=6$ to $60$, while the peak intensities show an effective scaling
close to $N^2$. The same pair has no crossing under independent decay,
identifying collective radiative kinetics as the origin of the effect.
\end{abstract}

\maketitle

\section{Introduction}
\label{sec:intro}

The Mpemba effect challenges the usual expectation that a system initially
closer to equilibrium should relax faster than one prepared farther away.
Originally associated with the observation that hotter water may freeze before
colder water \cite{MpembaOsborne1969}, it is now used more broadly for a
finite-time reversal of relaxation order. In Markovian dynamics, such a
reversal can result from the different weights with which initial states excite
slow decay modes \cite{LuRaz2017,KlichEtAl2019,kumar2020exponentially,joshi2024observing,ares2025quantum,zhang2025observation}.

Quantum versions have been investigated in both open and isolated systems.
Examples include accelerated stationarity after an optimized initial rotation
\cite{CarolloEtAl2021,moroder2024thermodynamics, caldas2026exponentially}, quantum dots coupled to reservoirs
\cite{ChatterjeeEtAl2023}, nonequilibrium open systems
\cite{NavaEgger2024,WangWang2024}, and symmetry restoration after many-body
quenches \cite{AresEtAl2023}. These works also emphasize that an anomalous
crossing can depend on the diagnostic: energy or current crossings are useful
witnesses, but a distance between states provides a more direct ordering with
respect to stationarity.

Here we ask the following question: can a Mpemba crossing occur while both initial
states undergo superradiantly enhanced emission? To answer this question, we consider $N$ two-level
emitters coupled to a common zero-temperature reservoir. In the symmetric
Dicke manifold \cite{Dicke1954,GrossHaroche1982}, the transition rate from the
level with $n$ excitations is proportional to $n(N-n+1)$. It is therefore only
of order $N$ at the boundaries of the ladder but reaches order $N^2$ near its
center. This nonuniformity makes relaxation depend on where the population is
placed, rather than only on its mean energy.

Collective models have previously appeared in the quantum Mpemba literature ~\cite{de2026collective, das2025mpemba,saliba2025unraveling}. However our proposal differs from all the previous ones:
we prepare sparse states diagonal in the Dicke basis and let both evolve under
the same uncontrolled collective decay. The mechanism is described directly
by positive population currents along the ladder, without assigning a
probabilistic meaning to individual Liouvillian modes.

We exhibit a family for which the initially more energetic and more distant
state crosses the reference state in both energy and trace distance. Both
states also display collective peaks above their independent-emitter
counterparts, and the effect persists from $N=6$ to $60$. Finally, we show
analytically that the same pair cannot cross under independent decay. We refer
to this coexistence as \emph{superradiant Mpemba relaxation}.

\section{Model and diagnostics}
\label{sec:model}

We consider $N$ identical two-level emitters with collective operators
$J_z=\frac12\sum_i\sigma_z^{(i)}$ and $J_-=\sum_i\sigma_-^{(i)}$. Choosing the
collective ground-state energy as zero, the Hamiltonian is
\begin{equation}
 H=\hbar\omega_0\left(J_z+\frac{N}{2}\right).
 \label{eq:H}
\end{equation}
Collective spontaneous emission into a common zero-temperature Markovian
reservoir is described by
\begin{equation}
 \dot\rho=-\frac{i}{\hbar}[H,\rho]+\Gamma\mathcal D[J_-]\rho,
 \qquad
 \mathcal D[L]\rho=L\rho L^\dagger-\frac12\{L^\dagger L,\rho\}.
 \label{eq:master}
\end{equation}
We restrict the dynamics to the fully symmetric sector $J=N/2$ and label the
Dicke states by their excitation number,
$|n\rangle\equiv|J=N/2,m=-N/2+n\rangle$. The ladder operator acts as
\begin{equation}
 J_-|n\rangle=\sqrt{n(N-n+1)}\,|n-1\rangle .
 \label{eq:ladder}
\end{equation}

For an initial state diagonal in this basis,
$\rho(0)=\sum_{n=0}^Np_n(0)|n\rangle\langle n|$, no Dicke-basis coherence is
generated. The populations obey the exact cascade
\begin{equation}
 \dot p_n=\Gamma(n+1)(N-n)p_{n+1}
 -\Gamma n(N-n+1)p_n,
 \label{eq:rates}
\end{equation}
with $p_{N+1}=0$. Hence the transition $|n\rangle\to|n-1\rangle$ occurs at
\begin{equation}
 \Gamma_n=\Gamma n(N-n+1).
 \label{eq:Gamma_n}
\end{equation}
The unique stationary state within the symmetric sector is
$\rho_{\mathrm{ss}}=|0\rangle\langle0|$. This uniqueness is restricted to the fully symmetric sector:
the full collective Liouvillian also supports dark stationary
states in lower-$J$ sectors.

The excitation energy and photon-emission rate are
\begin{align}
 E(t)&=\hbar\omega_0\sum_n n p_n(t),
 &
 I_{\mathrm{coll}}(t)&=\Gamma\sum_n n(N-n+1)p_n(t),
 \label{eq:observables}
\end{align}
and satisfy $\dot E=-\hbar\omega_0 I_{\mathrm{coll}}$. For independent
emitters with the same single-emitter rate, the corresponding intensity is
$I_{\mathrm{ind}}=\Gamma\langle n\rangle$ and its maximum occurs at $t=0$.
We identify collective enhancement when
$I_{\mathrm{coll}}^{\max}>I_{\mathrm{ind}}^{\max}$ for the same initial
population distribution.

Because all states considered below commute, the trace distance to stationarity
equals the total-variation distance of their populations. For the absorbing
ground state it reduces to
\begin{equation}
 D(t)\equiv\frac12\|\rho(t)-\rho_{\mathrm{ss}}\|_1
 =\DTV[\bm p(t),\bm p_{\mathrm{ss}}]=1-p_0(t).
 \label{eq:distance}
\end{equation}
We compare states $A$ and $B$ satisfying $E_A(0)>E_B(0)$ and
$D_A(0)>D_B(0)$. A distribution-level Mpemba crossing occurs at a finite
$t_D^\times$ after which $D_A(t)<D_B(t)$. We also record the energy crossing
$t_E^\times$ and the inversion depth
\begin{equation}
 \mathcal M_D=\max_t[D_B(t)-D_A(t)].
 \label{eq:depth}
\end{equation}

\section{A sparse superradiant Mpemba pair}
\label{sec:sparse}

For every $N$, define $n_A=\lceil2N/3\rceil$,where \(\lceil x\rceil\) denotes the smallest integer greater than or equal to \(x\), and prepare
\begin{equation}
 \rho_A(0)=|n_A\rangle\langle n_A|,
 \qquad
 \rho_B(0)=\frac12\left(|0\rangle\langle0|+|N\rangle\langle N|\right).
 \label{eq:pair}
\end{equation}

\begin{figure*}[!t]
 \centering
 \includegraphics[width=\textwidth]{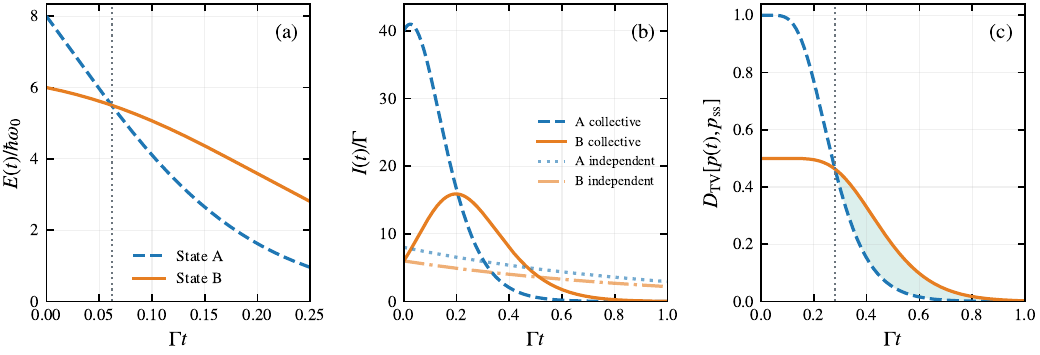}
 \caption{Superradiant Mpemba relaxation for $N=12$.
 (a) Excitation energies; the dotted line marks
 $\Gamma t_E^\times\simeq0.0617$.
 (b) Collective emission rates and independent-emitter references.
 (c) Trace distances to the stationary ground state; the dotted line marks
 $\Gamma t_D^\times\simeq0.2798$, and the shaded region indicates the reversed
 ordering $D_A<D_B$.}
 \label{fig:main}
\end{figure*}

These states are sparse and require no numerical optimization. Their initial
energies are $E_A(0)=\hbar\omega_0n_A$ and
$E_B(0)=\hbar\omega_0N/2$. Since $n_A>N/2$, state $A$ is more energetic.
Moreover, Eq.~\eqref{eq:distance} gives
\begin{equation}
 D_A(0)=1>D_B(0)=\frac12.
 \label{eq:initial_order}
\end{equation}

Figure~\ref{fig:main} shows the representative case $N=12$, for which
$\rho_A=|8\rangle\langle8|$ and
$\rho_B=(|0\rangle\langle0|+|12\rangle\langle12|)/2$. Although $A$ starts
both more energetic and farther from stationarity, the energy curves cross at
$\Gamma t_E^\times=0.0617$, followed by a trace-distance crossing at
$\Gamma t_D^\times=0.2798$. The maximum inversion is
$\mathcal M_D=0.1792$, so the effect is not a near-tangential numerical
intersection.

Both states exhibit collective enhancement. For $A$ and $B$, respectively,
\begin{equation}
 \frac{I_{\mathrm{coll}}^{\max}}{\Gamma}=(40.984,15.905),
 \qquad
 \frac{I_{\mathrm{ind}}^{\max}}{\Gamma}=(8,6).
 \label{eq:N12_peaks}
\end{equation}
State $A$ displays an almost immediate superradiant pulse, whereas the excited
component of $B$ develops a delayed burst.

The origin of this behavior is visible directly in the Dicke ladder. Writing
the downward probability current as
\begin{equation}
 \mathcal J_n(t)=\Gamma_n p_n(t),
 \label{eq:current}
\end{equation}
the intensity is $I_{\mathrm{coll}}=\sum_n\mathcal J_n$. For $N=12$, state
$A$ begins at $n=8$, where $\Gamma_8=40\Gamma$, immediately adjacent to the
maximum rates $\Gamma_6=\Gamma_7=42\Gamma$. The emitting half of state $B$
begins at the upper boundary, where $\Gamma_{12}=12\Gamma$; including its
weight $1/2$, its initial contribution is only $6\Gamma$. Thus
\begin{equation}
 I_A(0)=40\Gamma,
 \qquad
 I_B(0)=6\Gamma.
 \label{eq:initial_intensities}
\end{equation}

\begin{figure}[!t]
 \centering
 \includegraphics[width=\columnwidth]{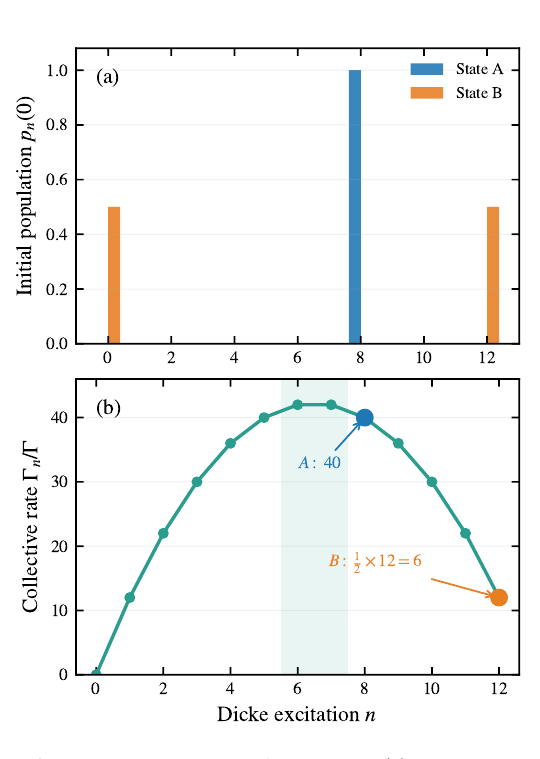}
 \caption{Kinetic mechanism for $N=12$.
 (a) Initial Dicke populations of the sparse pair.
 (b) Collective rate profile $\Gamma_n/\Gamma=n(13-n)$. State $A$ begins in a
 highly radiative level, whereas the emitting component of $B$ begins at the
 upper boundary. The labels include the initial population weights, giving
 $I_A(0)/\Gamma=40$ and $I_B(0)/\Gamma=6$.}
 \label{fig:mechanism}
\end{figure}

Energy responds to every downward transition, while the distance decreases
only when probability completes the last step $|1\rangle\to|0\rangle$, since
$-\dot D=N\Gamma p_1$. This explains why the energy crossing precedes the
distance crossing. The mechanism is therefore kinetic: $A$ enters the fast
portion of the cascade immediately, whereas the excited part of $B$ must first
descend from the boundary. No claim about a projection onto a unique fast
Liouvillian eigenvector is required.

\section{System-size dependence}
\label{sec:size}

We apply the same construction, without size-dependent optimization, to
\begin{equation*}
 N=6,8,10,12,16,20,24,30,40,50,60.
\end{equation*}

\begin{figure*}[!t]
 \centering
 \includegraphics[width=\textwidth]{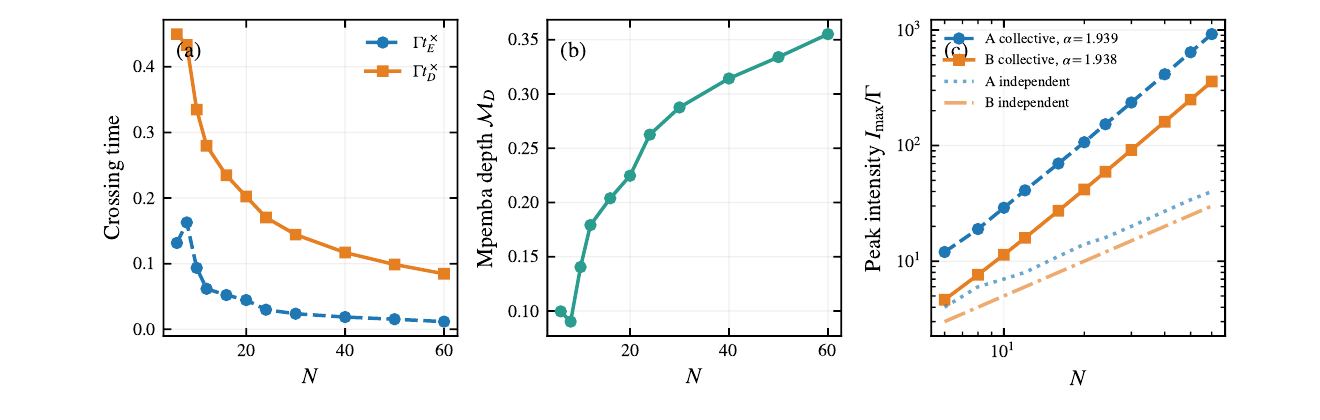}
 \caption{System-size dependence of the sparse family.
 (a) Energy- and trace-distance crossing times.
 (b) Maximum trace-distance inversion.
 (c) Collective peak intensities and independent-emitter references. Log--log
 fits for $N\ge12$ yield the finite-size exponents $1.939$ for $A$ and $1.938$
 for $B$.}
 \label{fig:size}
\end{figure*}

Both the energy and trace-distance crossings occur for every examined size, as
shown in Fig.~\ref{fig:size}(a). The crossing times decrease overall with $N$:
for example, $\Gamma t_D^\times$ changes from $0.4499$ at $N=6$ to $0.0846$
at $N=60$. Small nonmonotonic variations at low $N$ arise from the discrete
prescription $n_A=\lceil2N/3\rceil$. The inversion depth remains finite and
generally increases, reaching $\mathcal M_D=0.3552$ at $N=60$
[Fig.~\ref{fig:size}(b)].

The independent peaks scale linearly: $I_{A,\mathrm{ind}}^{\max}=\Gamma n_A$
and $I_{B,\mathrm{ind}}^{\max}=\Gamma N/2$. By contrast, log--log fits of the
collective peaks for $N\ge12$ give
\begin{equation}
 I_{A,\mathrm{coll}}^{\max}\propto N^{1.939},
 \qquad
 I_{B,\mathrm{coll}}^{\max}\propto N^{1.938},
 \label{eq:fits}
\end{equation}
with coefficients of determination above $0.9999$. These are finite-size
effective exponents, not universal critical exponents. They are consistent
with the characteristic quadratic scale of Dicke superradiance. For $A$, this
leading behavior is already explicit in
$I_A(0)=\Gamma n_A(N-n_A+1)=(2/9)\Gamma N^2+O(N\Gamma)$; for $B$, it emerges
as the fully excited component reaches the central ladder region.

\section{Discussion}
\label{sec:discussion}

The same sparse pair provides a simple test of whether collectivity merely
enhances an otherwise unrelated crossing. Under independent decay,
\begin{equation}
 E_A^{\mathrm{ind}}(t)=\hbar\omega_0n_Ae^{-\Gamma t}
 >\frac{\hbar\omega_0N}{2}e^{-\Gamma t}
 =E_B^{\mathrm{ind}}(t)
 \label{eq:no_energy_cross}
\end{equation}
at every finite time. There is no distance crossing either. Indeed, with
$x=1-e^{-\Gamma t}$,
\begin{equation}
 D_A^{\mathrm{ind}}=1-x^{n_A},
 \qquad
 D_B^{\mathrm{ind}}=\frac12(1-x^N).
 \label{eq:ind_distances}
\end{equation}
Since $2n_A>N$ and $0<x<1$,
$1-2x^{n_A}+x^N>(1-x^{n_A})^2\ge0$, implying
$D_A^{\mathrm{ind}}>D_B^{\mathrm{ind}}$. The collective factor
$n(N-n+1)$ therefore generates, rather than simply accompanies, the crossing
for this family.

Our result is complementary to modal approaches to the Mpemba effect. In
Ref.~\cite{CarolloEtAl2021}, a unitary rotation is optimized to suppress a slow
Liouvillian mode, producing a strong acceleration. Here no initial control
operation is applied, and we claim a finite-time reversal rather than exact
elimination of the slowest contribution. The positive currents
$\mathcal J_n=\Gamma_np_n$ provide a direct physical explanation without requiring a detailed mode-by-mode decomposition.

Although Eq.~\eqref{eq:rates} has the form of a classical pure-death process,
its nonlinear rates arise from the collective quantum matrix elements of
$J_-$. Moreover, the intermediate Dicke state $|n_A\rangle$ is generally
entangled in the local-emitter basis. The crossing does not, however, rely on
coherence between different Dicke levels and should not be presented as a
coherence or entanglement witness. It is most accurately described as a
population-level Mpemba effect generated by quantum cooperative kinetics.

The ideal model assumes permutation symmetry, zero temperature,
and negligible individual decay, dephasing, and leakage into
subradiant total-spin sectors.
Within any fixed lower-$J$ sector, collective decay generates an
analogous Dicke cascade, whereas extending the present Mpemba
construction to states with support on different total-spin
sectors is nontrivial because the full collective dynamics
possesses a degenerate dark manifold.
These assumptions are appropriate for demonstrating the mechanism,
but they also delimit the present claim.
The sparse preparation is operationally simple in principle:
$A$ is a fixed-excitation Dicke state, while $B$ can be realized
by randomizing equally between the collective ground and fully
excited states.
Experimentally, the energy follows from the mean excitation number,
the superradiant signal from the photon flux, and the trace distance
from the ground-state probability through $D=1-p_0$.

\section{Conclusion}
\label{sec:conclusion}

We have presented a simple family of Dicke states for which an initially more
energetic and more distant state overtakes a closer one during collective
spontaneous emission. The reversal occurs in both energy and trace distance,
while both states develop emission peaks above their independent-emitter
references. The mechanism is the nonuniform Dicke rate profile: the farther
state starts directly in a highly radiative region, whereas the emitting part
of the closer state begins at the upper boundary. The crossing persists from
$N=6$ to $60$, and the collective peaks approach the quadratic superradiant
scale. Because the same pair cannot cross under independent decay, collective
radiative kinetics is essential to the effect. These results provide a compact
proof of principle that Mpemba relaxation and superradiance can occur jointly
in an open many-body quantum system.

\vspace{0.2cm}

\appendix
\section{Numerical details}
\label{app:numerics}

Writing $\tau=\Gamma t$ and $\mathsf K=\mathsf M/\Gamma$, the populations were
propagated as $\bm p(\tau)=e^{\mathsf K\tau}\bm p(0)$ by applying the matrix
exponential directly to the initial vector, without diagonalizing
$\mathsf K$. Production data used 4001 uniformly spaced points on
$0\le\tau\le1.5$. A crossing was located from a sign change of
$X_A-X_B$ between adjacent points and refined by linear interpolation. Values
smaller than $10^{-14}$ were set to zero. Repeating the calculations with 2001
and 8001 points changed the reported crossing times and peak intensities by at
most a few parts in $10^6$ under the final refinement. The log--log intensity
fits used ordinary least squares over the points with $N\ge12$.

\begin{acknowledgments}
\end{acknowledgments}
The authors acknowledge financial support from CAPES, CNPq, and FAPESP, Brazilian funding agencies.

\bibliography{References.bib}

\end{document}